\documentclass[12pt,preprint]{aastex}

\slugcomment{Submitted to \apj Letters}

\shorttitle{RXTE Absolute Timing of Crab Pulsar}
\shortauthors{Rots et al.}

\begin{document}


\title{Absolute Timing of the Crab Pulsar with RXTE}


\author{Arnold H. Rots}
\affil{Harvard-Smithsonian Center for Astrophysics, 60 Garden
Street MS 67, Cambridge, MA 02138}
\email{arots@head.cfa.harvard.edu}
\author{Keith Jahoda}
\affil{Laboratory for High Energy Astrophysics, Code 660, NASA,
Goddard Space Flight Center, Greenbelt, MD 20771}
\and
\author{Andrew G. Lyne}
\affil{Department of Physics and Astronomy, University of Manchester, Jodrell Bank,
Macclesfield, SK11 9DL, UK}



\begin{abstract}
We have monitored the phase of the main X-ray pulse of 
the Crab pulsar with the Rossi X-ray Timing Explorer (RXTE) for almost
eight years, since the start of the mission in January 1996. The
absolute time of RXTE's clock is sufficiently accurate to allow this
phase to be compared directly with the radio profile.

Our monitoring observations of the pulsar took place bi-weekly (during the periods when 
it was at least 30 degrees from the Sun) and we correlated the data with radio 
timing ephemerides derived from observations made at Jodrell Bank.
We have determined the phase of the X-ray main pulse for each
observation with a typical error in the individual data points of
50~$\mu$s.  The total ensemble is consistent with a phase that is
constant over the monitoring period, with the X-ray pulse leading the
radio pulse by 0.0102$\pm$0.0012 period in phase, or 
344$\pm$40~$\mu$s in time.  The error estimate is dominated by a
systematic error of 40~$\mu$s in the radio data, arising from
uncertainties in the variable amount of pulse delay due to
interstellar scattering and instrumental calibration.  The
statistical error is 0.00015 period, or 5~$\mu$s.
The separation of the main pulse and interpulse
appears to be unchanging at time scales of a year or less, with an
average value of 0.4001$\pm$0.0002 period.  There is
no apparent variation in these values with energy over the 2-30 keV range.

The lag between the radio and X-ray pulses may be constant in phase
(i.e., rotational in nature) or constant in time (i.e., due to a
pathlength difference).  We are not (yet) able to distinguish between
these two interpretations.
\end{abstract}




\section{Introduction}

For many years it has been assumed that the main pulse and interpulse
of the Crab pulsar (PSR~B0531+21) are perfectly lined up in
phase over the full range of the electro-magnetic spectrum.
Even though there have been reports in the past that this alignment
may not be as perfect as generally assumed, the absolute calibration
of spacecraft clocks was not sufficiently accurate to allow a precise
measurement of the phase difference.
The most compelling result predating the Rossi X-ray Timing Explorer
(RXTE) observations was presented by \citet{masn1994}, based on
Figaro~II observations covering 0.15 to 4.0~MeV, that were made in 1986 and
1990.  While discounting the 1986 result which has a considerable
uncertainty due to potential errors in the dispersion measure, we
consider the 1990 result (the $\gamma$-ray pulse leading the radio
pulse by 375$\pm$148~$\mu$s) fairly reliable, though the error
is probably underestimated.

The precision with which absolute time can be determined with the RXTE
clock allows
us to time X-ray pulses with an accuracy better than 10~$\mu$s,
depending on pulse shape, as shown by \citet{rots1998b}. At the same time, 
the Crab pulsar monitoring program at Jodrell Bank provides timing
ephemeris data, reduced to infinite frequency.
These (monthly) timing ephemerides represent fits to the daily
time-of-arrival measurements with rms residuals of order 20-50~$\mu$s.
This allows us to measure and monitor the radio to X-ray phase
difference of the pulses with an
error of about one milli-period. We have reported on these results
in the past (\citet{rots1998a}, \citet{rots1998c}, \citet{rots2000a},
\citet{rots2000b}).

At optical wavelengths, \citet{sanw1999} has reported a time delay of
140~$\mu$s (optical leading the radio), but the details are not easily
accessible.  \citet{shea2003} report that, in the case of giant radio
pulses, the optical pulse in the wavelength range 600-750~nm is
leading the radio pulse by 100$\pm$20~$\mu$s.
\citet{roma2001}, on the other hand, claim that the optical
(355-825~nm) and radio peaks are coincident within 30~$\mu$s, based on
test observations with a prototype transition-edge sensor detector.
However, it is not clear whether the timing calibration of the
instrument was complete at the time.

\citet{ulme1994} presented results from OSSE observations
(50-100~keV), indicating that the hard X-ray to radio lag was
$<$30$\pm$30~$\mu$s.  It would appear that the estimate of their
errors was too optimistic. \citet{nola1993} present pulse profiles but
no absolute phases. \citet{kuip2003} report on INTEGRAL data, covering
6-50~keV, and deriving a time delay (the radio trailing) of
280$\pm$40~$\mu$s for a single epoch.

The precise timing of the pulses in the different wavelength regimes
has important repercussions for the understanding of the nature and
spatial origin of the emission processes that give rise to the pulses
in different parts of the spectrum.  \citet{roma1995} have suggested
that, while the radio precursor comes from the polar cap, the pulse
and interpulse originate in the outer gap in the magnetosphere, with
the higher energy pulses being generated at significantly greater
height.  Thus, measuring the pulse shapes and the absolute timing
throughout the electromagnetic spectrum places important constraints
on the shape of the the outer gap and on the height in the
magnetosphere at which the radiation is generated.  

In this paper we will present the results
of the RXTE monitoring campaign of the Crab Pulsar from the start of
that mission.  We have adopted the radio nomenclature
for the features in the pulse profile (main pulse, bridge, and interpulse).

\section{Observations}

The observations presented here were made as part of an on-going monitoring 
campaign of the Rossi X-ray Timing Explorer's (RXTE). Observations
of about 1000~s in duration were initially made at 
weekly, later bi-weekly, intervals, with the exception of a period
from mid-May until mid-July when the Crab pulsar is too close to the sun. 

In this paper we shall use the events in the 2-16 keV range, collected
with the RXTE's Proportional Counter Array (PCA) in 177
observations between MJD 50129 and 52941.  For the first four and a
half years the observations were made in pulsar fold mode, with
approximately 80 bins per period.  Halfway through the fifth year this
was changed to an event mode with 250~$\mu$s time resolution.

The accuracy of the RXTE clock in absolute time is about 8~$\mu$s for
data taken before 29 April 1997 (MJD 50567); see
\citet{rots1998b}. After that date, the error decreased to 2~$\mu$s
\citep{mark2003}.
This accuracy can be achieved by applying the fine clock corrections
supplied by the RXTE GOF.  In addition to the clock correction, there
is an instrumental delay correction for the PCA of 16-20~$\mu$s.
Without the fine clock correction the uncertainty in absolute time is
100~$\mu$s.

The radio timing ephemeris is derived from Jodrell Bank observations,
daily at 610~MHz and weekly at 1420~MHz.  See, e.g., \citet{lyne1993}.
Reduced to ``infinite frequency'', the ephemeris provides the
dispersion-corrected time
of arrival of the center of the main pulse and is published on the
world-wide web\footnote{\url{http://www.jb.man.ac.uk/$\sim$pulsar/crab.html}}. The ephemeris records contain:
range of validity (MJD, UTC); phase-zero in MJD (UTC; geocenter);
RA, Dec (J2000, FK5); $\nu$ and its first two derivatives; rms of the solution's fit.
The timing ephemeris records, each of which covers one month, are
created by the Tempo package, on the basis of the JPL solar system
ephemeris DE200.
In the period prior to MJD 50870 the Crab pulsar suffered a
substantial amount of variable multipath scattering within the nebula;
see also \citet{wong2001}.
While the Jodrell Bank staff endeavored to remove the effects of this
from the data, there was uncertainty in doing so, and this is
reflected in the quoted errors and in the increased scatter.
Apart from uncertainty in the delay due to
scattering, which may change on timescales of a few months, the errors
quoted in the ephemeris also contain a contribution arising from
unknown systematic effects in the system, such as unmodeled delays in
filterbanks and imperfect polarization calibration.  This amounts to about
40~$\mu$s and should not be treated as a statistical error which
reduces in a known fashion upon averaging.

\section{Analysis}

\begin{figure}
\plotone{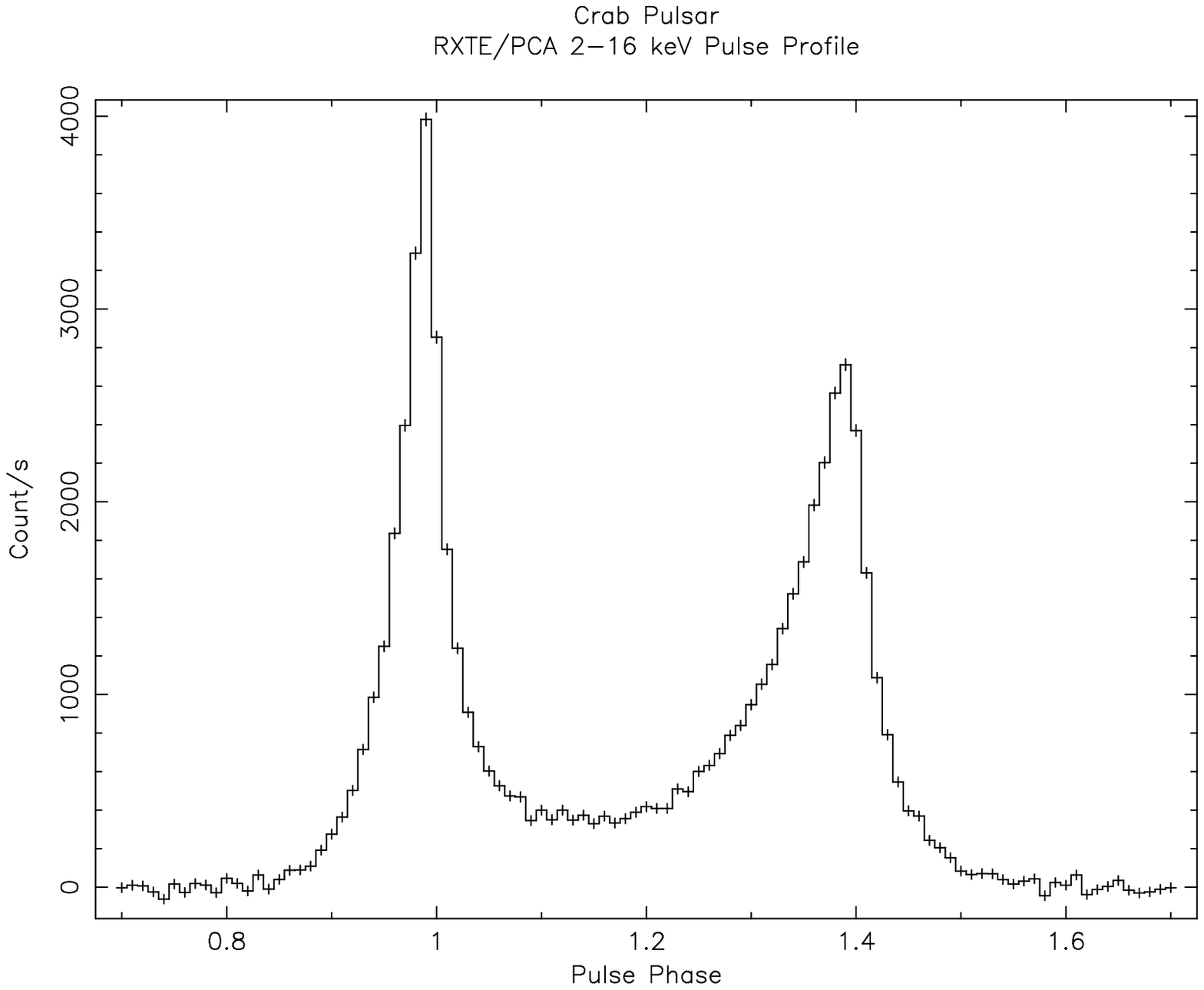}
\caption{Typical RXTE pulse profile for the Crab pulsar.  The events
are binned in 100 bins per period; errors are indicated by vertical
bars. \label{fig1}}
\end{figure}

The observations were analyzed using the program {\em faseBin} which applies the 
barycenter correction, ties the arrival times to the radio timing ephemeris, and bins 
the events in absolute phase. {\em faseBin} is the core of the {\em
Ftool} \citep{black1995} {\em fasebin}.
The unpulsed component is then subtracted and the data are 
integrated over the energy range 2-16 keV.
A typical 2-16 keV pulse profile is shown in Fig.~\ref{fig1}.

\notetoeditor{Please insert Fig. 1}

At this point we needed to decide how to define the phase of the main
X-ray pulse.  The problem is that, while the radio pulse is very
symmetric, the X-ray pulse is clearly asymmetric.  In order to avoid
any assumptions concerning the modeling of the pulse shape to enter
into our analysis, we decided to use the peak of the pulse as
representative of the X-ray phase.  To this end we used three
different peak finding algorithms, each designed to be free of model
assumptions to the extent possible: a parabolic fit to the highest bin
in the profile and its two neighbors, using 200 phase bins; fitting a
Lorentzian function to the phase range 0.98 to 1.00, using 400 phase
bins; and calculating the first moment over that part of the pulse
where the bins exceed 80\% of the highest bin, using 800 phase bins.
This procedure allows us to determine the pulse phase in an individual
observation with an accuracy of 1~milliperiod.  We did investigate
whether the time resolution of the observations gives rise to
an additional systematic error and found this not to be the case:
Lorentzian fits to two back-to-back observations made in the fall of
2003 with resolutions of 250~$\mu$s and 16~$\mu$s, respectively,
differed by less than 0.1 milliperiod.

The absolute phase of the peak of the X-ray main pulse (with respect
to the peak of the radio main pulse, using the Lorentzian fits), as a 
function of time (in Modified Julian Days) is shown in
Fig.~\ref{fig2}; the errors are 
a combination of the 1~milliperiod error mentioned above and the rms
deviations in the fits of the radio timing ephemerides and typically
amount to 40~$\mu$s.  Independently from these statistical errors
there is the systematic error of up to 40~$\mu$s introduced by the
radio receiver system and calibration, as mentioned in the previous
section.

\notetoeditor{Please insert Fig. 2; if possible over the full page
width; however, if that leads to exceeding the 4-page limit, then one
column width will do.}

\section{Results and Discussion}

\begin{figure}
\plotone{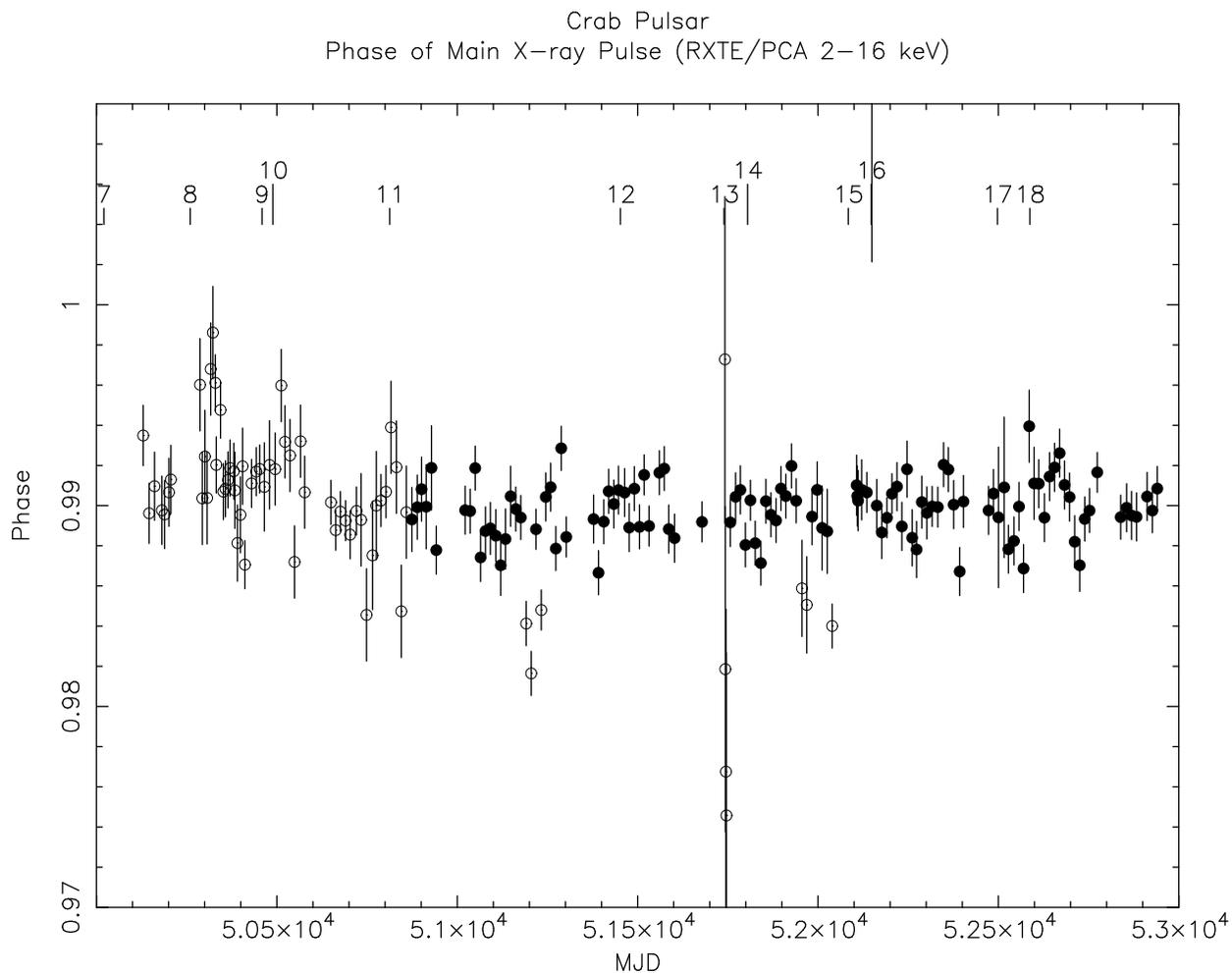}
\caption{Phase of the X-ray main pulse relative to the Jodrell Bank
radio timing ephemerides for 177 RXTE observations as a function of
time. The phase is calculated by fitting a Lorentzian function to the
phase range 0.98 to 1.00 with 0.0025 bin size (see text). The error
bars represent statistical errors only.  Filled
circles represent high-quality data points, open circles
data points of questionable quality. Numbered tick marks indicate the
times of glitches (see text).
\label{fig2}}
\end{figure}

Fig.~\ref{fig2} shows the history of the main X-ray pulse phase, with the
occurrences of glitches marked. Glitches 7 through 12 are taken from
\citet{wong2001}, while the glitch numbered 8 corresponds to Note 12
on the Jodrell Bank web page and adding one to glitch 
numbers 12 through 18 yields the corresponding Note numbers (13
through 19) on that page. We have monitored the X-ray emission more
closely following two glitches, but found no unusual behavior.

The data points in this figure can be divided into four quality
categories.  First, the 
data points that are based on timing ephemerides prior to MJD 50870
obviously display larger deviations than the later ones; we attribute
this to the poorer quality of those radio ephemeris records.  Second,
there are a number of points with high error estimates
($>$5~milliperiods), associated with glitches; the monthly timing
ephemeris records are not sufficiently fine-grained to handle glitches
properly.  Third, there are seven
outlyers (below phase 0.9860) that are clearly well below the remaining
data points. In all cases these represent either all observations
covered by a single ephemeris record or observations at the edge of
such a record; hence, we attribute these to inaccuracies in the timing
ephemerides. Fourth, the remaining 111 data points form a normal
distribution with the expected rms scatter of 1~milliperiod. This
indicates that the data are consistent with a constant value.  The data
points from the first three categories are represented by open circles
in Fig.~\ref{fig2}; the high-quality data points (category 4) are shown as
filled circles.

In order to determine the phase of the X-ray main pulse we have
excluded all data points that were deemed flawed (i.e., in the first
three categories above). Least-squares fits (weighted averages) to the
results from the three different peak-finding algorithms that we used
lead us to conclude that the X-ray main pulse leads the radio main
pulse (as defined by the radio timing ephemerides) by
10.25$\pm$0.15~milliperiod, or 344$\pm$5~$\mu$s, with a
reduced $\chi^{2}$ of 1.3.  The quoted errors represent the
differences between the results from the three methods.  The
statistical errors in the three individual fits are smaller.
In addition, of course, there is still the uncertainty of the
40~$\mu$s systematic error in the radio ephemerides.
We emphasize that, although we believe these error estimates to be
realistic, a different definition of the pulse phase may lead to
larger discrepancies.  Ideally, one should analyze the data that are
available in the different spectral bands with a uniform pulse
definition.

The result obtained by \citet{kuip2003} of 280$\pm$40~$\mu$s for
a single INTEGRAL observation is probably to be considered consistent
with our findings, especially since it used a timing ephemeris record
at MJD 52685 that gives rise to slightly elevated phase values in our
data.  However, the phases that they quote for
the main pulse on MJD 52683 and MJD 52697, derived from the same RXTE
observations that we have used, differ from our values by
+1.7 and +1.1 milliperiods, respectively.  We believe that this
difference is due to the definition of the phase that is used by these
authors. \citet{kuip2003} define the phase of the main peak as the
position of an asymmetric Lorentzian fit to the phase range 0.95 to
1.05.  This definition, in our opinion, is not as free of
model-dependent assumptions as our analysis methodology; it appears
that there is a systematic offset of about 40-50~$\mu$s.  Note that,
since these authors used the same Jodrell Bank timing ephemeris
records, the radio systematic error does not play a role here.

As to the question whether the lag between the X-ray and radio pulses
is constant in phase (i.e., the lag is rotational in nature)
or in time (i.e., the lag represents a pathlength difference), the
data are not conclusive.  The 
former would require the phase offset in Fig.~\ref{fig2} to be constant with
time, while the latter would require the phase to increase
linearly with a slope of $+1.0\times10^{-8}$ period/day. A linear fit
to the good data in Fig.~\ref{fig2} yields a slope of
$(+3.3\pm2.0)\times10^{-7}$.  This result is probably affected by
Malmquist bias and possibly other sources of systematic errors.  Unless
our measurement accuracy can be dramatically improved, it will require at
least another seven years of monitoring before we can answer this
question definitively in this manner.  If indeed we are dealing with a
time offset, this would correspond to a pathlength difference of
about 100~km.

Additional analysis of the RXTE data reveals that the PCA and HEXTE
pulses are perfectly aligned to within 1~milliperiod (i.e., no phase
change over the 2 to 30~keV energy range), while pulse phase
determinations over a 12-hour Crab observation show less than
1~milliperiod variation in the phase of the main pulse.  This is all
within the measurement errors.

We have measured the phase
difference between the X-ray main pulse and interpulse, a quantity
that is independent of any uncertainties in the radio timing ephemeris
records. The average value over the 7.6 year period is
0.4001$\pm$0.0002 period.  There do not appear to be any variations on
time scales of the order of a year or less, but we cannot entirely
exclude systematic variations or systematic errors of the order of
1~milliperiod on time scales of several years.

\section{Conclusion and Summary}
The X-ray main pulse leads its radio counterpart by about
344$\pm$40~$\mu$s (systematic error); the statistical error is
5~$\mu$s.  This is more than twice the time difference of 
140~$\mu$s that 
\citet{sanw1999} determined for the optical B band and three times the
100~$\mu$s measured by \citet{shea2003} in red light.
The time or phase difference appears to have been
constant over the past eight years, but the data are not accurate
enough to distinguish between the two.
The errors in individual measurements range up to 50~$\mu$s.
We should caution the reader when making comparisons with results in
other wave bands.  First, various authors have used different
definitions of the pulse phase.  In our estimation, our own definition
agrees with that used in the radio band, as does the definition of
\citet{shea2003}, but that is probably not true for most other
reports.  Second, the scatter in values for individual observations is
fairly large (a milliperiod) and may be intrinsic.  Greater accuracy
can only be achieved with a statistically significant set of
observations.

As we have mentioned, there is a systematic error of up to 40~$\mu$s
in the offset due to uncertainties in the interstellar scattering and the
calibration of the radio equipment.  This error may change on
timescales of a few months, but since we do not know whether (and if
so, by how much) the error is reduced by averaging, we quote 40~$\mu$s
as the final uncertainty for the radio to X-ray timing of the pulse.
However, such an error does not affect the comparison with results
from other wavebands provided they are all approximately
contemporaneous and use the same Jodrell Bank timing ephemeris
records.  On the other hand, it also appears that most (if not all)
errors for the results at other wavebands quoted in the Introduction
are seriously underestimated by ignoring the systematic error.

The phase difference between the two X-ray pulses is constant at 0.400
period, within the measurement errors. It is also equal to the phase
difference between the radio main pulse and interpulse, within the
measurement error.  It may be of interest to note that in the X-ray
pulse profile the trailing edges of the pulse as well as the
interpulse are distinctly steeper than their leading edges.  This does
not appear to be the case for the optical interpulse.

If the X-ray to radio lag were a true phase lag, attributable to the
(radial) energy distribution across a cone, with the pulses occurring
near the cone edges, one would expect the placement to be symmetrical,
i.e., one X-ray pulse to be leading, the other trailing.
As it stands, both X-ray pulses are leading by the same amount.
The simplest explanation for this phenomenon is that we are dealing
with a time delay reflecting a pathlength difference: the radio pulses
originate approximately 100~km closer to the surface of the neutron
star, as already suggested by \citet{masn1994}.

\acknowledgements
We gratefully acknowledge the efforts of Edward Morgan and Robert
Shirey to keep up with constructing pulsar fold mode configurations as
the pulsar slowed down.  We are indebted to the RXTE-GOF for
maintaining the RXTE fine clock correction file, tdc.dat.  We are
grateful to Robert Pritchard and Mark Roberts for maintaining the
Jodrell Bank Crab timing ephemerides.  We thank
Dr.\ Kuiper for helpful discussions on pulse fitting and an anonymous
referee for stimulating comments.
This research has been supported by NASA grant NAG5-7346 and
NASA contract NAS 8-39073 (CXC).

{}


\onecolumn

\end{document}